\documentclass[sigconf]{acmart}
\usepackage{soul}
\usepackage{graphicx}
\usepackage{xcolor}
\usepackage{tikz}
\usepackage{fontawesome5}
\usepackage{adjustbox}
\usepackage{multirow}
\usepackage{placeins}
\usepackage{enumitem}
\usepackage{balance}
\setlist[itemize]{leftmargin=*}

\AtBeginDocument{}

\setcopyright{none}
\acmConference[WWW '25]{Proceedings of the ACM Web Conference 2025}{April 28-May 2, 2025}{Sydney, NSW, Australia}
\acmBooktitle{Proceedings of the ACM Web Conference 2025 (WWW '25), April 28-May 2, 2025, Sydney, NSW, Australia}
\acmDOI{10.1145/3696410.3714507}
\acmISBN{979-8-4007-1274-6/25/04}

\newcommand{\subr}[1]{{\small\texttt{r/#1}}}

\definecolor{none}{HTML}{e6e6e6}
\definecolor{adaptation}{HTML}{fdbf6f}
\definecolor{adaptation-header}{HTML}{ff7f00}
\definecolor{personalization}{HTML}{a6cee3}
\definecolor{personalization-header}{HTML}{1f78b4}
\newcommand{\Ba}{\colorbox{none}{\texttt{Ba}}}
\newcommand{\Mu}{\colorbox{none}{\texttt{Mu}}}
\newcommand{\Hs}{\colorbox{none}{\texttt{Hs}}}
\newcommand{\Redd}{\colorbox{adaptation}{\texttt{Re}}}
\newcommand{\Prev}{\colorbox{adaptation}{\texttt{Pr}}}
\newcommand{\Summ}{\colorbox{personalization}{\texttt{Su}}}
\newcommand{\Hist}{\colorbox{personalization}{\texttt{Hi}}}

\newcommand*\emptydot[1][0.66ex]{\tikz\draw (0,0) circle (#1);} 
\newcommand*\fulldot[1][0.66ex]{\tikz\fill (0,0) circle (#1);}

\begin{document}

\title[Contextualized Counterspeech: Strategies for Adaptation, Personalization, and Evaluation]{Contextualized Counterspeech:\\Strategies for Adaptation, Personalization, and Evaluation}
\titlenote{\textcolor{red}{Article published in \textit{WebConf'25 – 34th ACM Web Conference}. DOI: \href{http://doi.org/10.1145/3696410.3714507}{10.1145/3696410.3714507}. Please, cite the published version.}}

\author{Lorenzo Cima}
\authornote{equal contributions}
\email{lorenzo.cima@phd.unipi.it}
\affiliation{\institution{University of Pisa and IIT-CNR}
  \country{Pisa, Italy}
}

\author{Alessio Miaschi}
\authornotemark[2]
\email{alessio.miaschi@ilc.cnr.it}
\affiliation{\institution{ILC-CNR}
  \country{Pisa, Italy}
}

\author{Amaury Trujillo}
\email{amaury.trujillo@iit.cnr.it}
\affiliation{\institution{IIT-CNR}
  \country{Pisa, Italy}
}

\author{Marco Avvenuti}
\email{marco.avvenuti@unipi.it}
\affiliation{\institution{University of Pisa}
  \country{Pisa, Italy}
}

\author{Felice Dell'Orletta}
\email{felice.dellorletta@ilc.cnr.it}
\affiliation{\institution{ILC-CNR}
  \country{Pisa, Italy}
}

\author{Stefano Cresci}
\email{stefano.cresci@iit.cnr.it}
\affiliation{\institution{IIT-CNR}
  \country{Pisa, Italy}
}

\renewcommand{\shortauthors}{Lorenzo Cima et al.}

\begin{abstract}
AI-generated counterspeech offers a promising and scalable strategy to curb online toxicity through direct replies that promote civil discourse. However, current counterspeech is one-size-fits-all, lacking adaptation to the moderation context and the users involved. We propose and evaluate multiple strategies for generating tailored counterspeech that is adapted to the moderation context and personalized for the moderated user. We instruct a LLaMA2-13B model to generate counterspeech, experimenting with various configurations based on different contextual information and fine-tuning strategies. We identify the configurations that generate persuasive counterspeech through a combination of quantitative indicators and human evaluations collected via a pre-registered mixed-design crowdsourcing experiment. Results show that contextualized counterspeech can significantly outperform state-of-the-art generic counterspeech in adequacy and persuasiveness, without compromising other characteristics. Our findings also reveal a poor correlation between quantitative indicators and human evaluations, suggesting that these methods assess different aspects and highlighting the need for nuanced evaluation methodologies. The effectiveness of contextualized AI-generated counterspeech and the divergence between human and algorithmic evaluations underscore the importance of increased human-AI collaboration in content moderation.

\noindent {\small\faExclamationTriangle}~~\textbf{Warning:} \textit{This paper contains examples that may be perceived as offensive or upsetting. Reader discretion is advised.}
\end{abstract}



\maketitle

\section{Introduction}
\label{sec:introduction}
Online toxicity refers to hateful, offensive, or otherwise harmful speech on the Web that can cause distress in readers or lead someone to abandon a conversation~\cite{wulczyn2017ex}. Toxicity has dire social and economic costs. Online, it reduces user participation, hinders information exchange, and deepens divides~\cite{aleksandric2024users}. Offline, it may lead to physical violence, reduce norm adherence, and cause severe psychological distress~\cite{gallacher2021online,pascual2021toxicity,saha2019prevalence}. Hence, it is a growing concern for both regulators and platform administrators~\cite{eu2019CoC}.

\begin{figure}[t]
\centering
    \includegraphics[width=1\columnwidth]{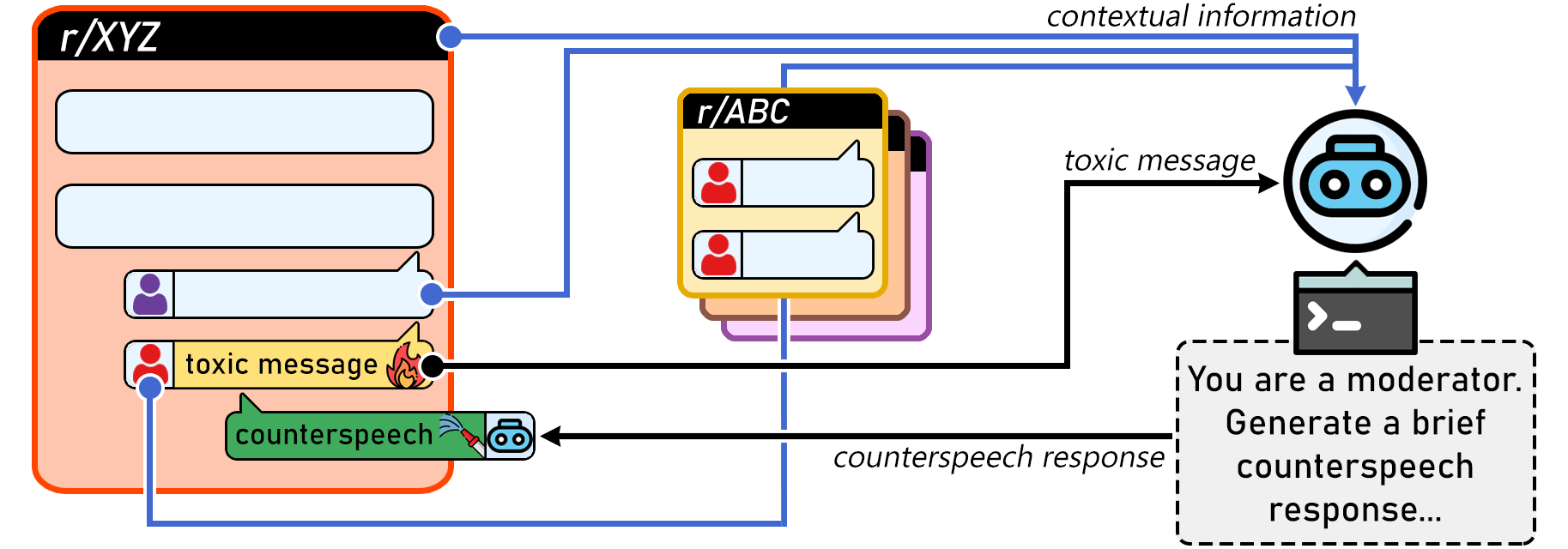}
    \caption{Current AI-generated counterspeech only leverages the content of the toxic message. Here, we generate contextualized counterspeech that also leverages information about the community, the conversation, and the moderated user to craft more persuasive responses.}
    \label{fig:intro-example}
\end{figure}

Platforms apply a wide array of moderation actions to curb toxicity and other online harms~\cite{trujillo2025dsa,dwork2024content}. These include content and user removals, friction interventions, and demotion~\cite{trujillo2022make,chandrasekharan2022quarantined,tessa2024beyond}. An alternative to centralized moderation is social correction---such as counterspeech---where users proactively respond to toxic content encouraging respectful and constructive communication~\cite{garland2022impact,bonaldi2024nlp}. Counterspeech holds promise as it avoids censorship concerns while preventing toxic users from merely relocating to other communities~\cite{horta2021platform}. However, a number of challenges currently limit its efficacy and applicability. The emotional toll on responders, who spend countless hours confronting hate, incivility, and personal attacks, is a major issue~\cite{roberts2019behind,steiger2021psychological}. Safety is another concern, as responders may face retaliation~\cite{doerfler2021m,tabassum2024investigating}. Moreover, the widespread nature of online toxicity makes manual counterspeech highly impractical. To overcome these limitations, automated counterspeech systems leveraging generative AI technologies---such as large language models (LLMs)---have been developed~\cite{tekirouglu2020generating,bonaldi2024nlp}. Yet, evaluating their effectiveness remains challenging~\cite{gillespie2020content}. Existing work primarily focused on basic characteristics like grammaticality and relevance, using quantitative indicators that offer a narrow assessment of effectiveness~\cite{zubiaga2024llm}. Additionally, current machine-generated counterspeech is one-size-fits-all~\cite{cresci2022personalized}, relying solely on the toxic message as input, which is often insufficient for crafting effective responses. On the contrary, effective moderation must be context-dependent~\cite{gillespie2020content,huang2024opportunities}, requiring a shift from generic to tailored approaches that consider the broader conversational context~\cite{cresci2022personalized}. 

\noindent\textbf{Research focus.} We propose using an LLM to generate \textit{adapted} and \textit{personalized} counterspeech messages against online toxicity. Our core novelty is the generation of contextualized counterspeech, as opposed to generic one. Our model ingests information about the community where the moderation occurs, the moderated user, and the specific conversation, as outlined in Figure~\ref{fig:intro-example}. We test the hypothesis that contextualized counterspeech can be more persuasive than generic one, in a set of political communities on Reddit. We also assess which strategies for adaptation and personalization are more effective. To test the hypothesis, we define the desired characteristics of effective counterspeech and we use quantitative indicators to measure them. For attributes that indicators cannot reliably capture~\cite{zubiaga2024llm}, we run a pre-registered, mixed design crowdsourcing experiment. Our results surface the challenges of generating contextualized counterspeech, but also show that contextual counterspeech can outperform state-of-the-art generic counterspeech in various aspects. In summary, our main contributions are:
\begin{itemize}
    \item We propose, experiment with, and evaluate strategies for generating counterspeech that is \textit{adapted} to the community and moderation context, and \textit{personalized} to the moderated user.
\item We assess the contribution that different types of contextual information provide, showing that contextual counterspeech can outperform generic one in terms of \textit{adequacy} and \textit{persuasiveness}, without compromising other characteristics.
    \item We show that evaluations derived from quantitative indicators strongly differ from human judgments, which bears implications for the development of counterspeech evaluation methodologies.
\end{itemize}
 \section{Related Work}
\label{sec:related_work}

\subsection{LLM-generated counterspeech}
Counterspeech can be more effective than other moderation actions, such as content and user removals, without limiting free speech~\cite{chung2023understanding}. Favorable results were obtained by scholars~\cite{hangartner2021empathy,bilewicz2021artificial}, NGOs~\cite{chung2019conan,chung2021towards}, and ordinary users~\cite{goffredo2022counter,hassan2023discgen} alike, as part of observational~\cite{garland2022impact}, quasi-experimental~\cite{bilewicz2021artificial}, and experimental~\cite{munger2017tweetment,hangartner2021empathy} studies. These positive results also extend to AI-generated counterspeech. However, we currently lack a thorough understanding of the types of counterspeech that are most effective and the optimal conditions for their generation~\cite{chung2023understanding}. For these reasons, research is focusing on identifying the conditions, methods, and resources that make AI-generated counterspeech effective. \citet{tekiroglu2022using} compared pre-trained LLMs and decoding strategies to highlight those capable of generating effective counterspeech. Multiple studies leveraged socio-psychological theories to generate messages based on different strategies, such as empathy, abstract norms, disapproval, and humor~\cite{sue2019disarming,hangartner2021empathy,bilewicz2021artificial,bar2024generative}. Others experimented with Retrieval-Augmented Generation (RAG) to integrate external knowledge flexibly and improve the specificity of the generated counterspeech~\cite{jiang2025rezg}. However, all these strategies were applied independently of the moderated context and user. As shown in Appendix Table~\ref{tab:related-work}, the only exceptions are \citet{dougancc2023generic} who used the age and gender of hateful users to generate tailored counterspeech, and~\citet{bar2024generative} who prompted an LLM to generate contextualized counterspeech based on the content of the toxic message. Herein, we go beyond existing work by generating adapted and personalized counterspeech that leverage content about the user, the conversation, and the community, assessing which type of contextual information results in effective counterspeech.

\subsection{Persuasiveness of LLM-generated messages}
LLMs persuasiveness has been explored from various perspectives. While some studies focused on persuasion directed at other LLMs~\cite{breum2024persuasive, salvi2024conversational}, most research examined their influence on humans. For instance,~\cite{furumai2024zero,shi2020effects} explored attempts to secure donations through different communication strategies, while~\cite{ramani2024persuasion} used LLMs as sales agents. Propaganda was the focus of~\cite{goldstein2024persuasive}, which demonstrated that AI-generated and human-generated propaganda exhibit comparable levels of persuasiveness, suggesting they can be used interchangeably. Similar results were obtained about political persuasiveness~\cite{hackenburg2024evidence}. Instead, \citet{costello2024durably} instructed an LLM to debunk conspiracy theories, achieving notable long-term effects, with persuasion lasting up to two months post-intervention. Finally, several studies examined the use of LLMs as a moderation tool for mitigating conflicts and for reducing toxic messages on social media~\cite{cho2023can,hong2024outcome,govers2024ai}. In spite of these results however, the persuasiveness of AI-generated contextualized counterspeech is still unexplored.

\subsection{LLM adaptation and personalization}
The majority of existing content moderation interventions is one-size-fits-all, where the intervention is applied uniformly to all moderated users. However, multiple studies have highlighted the limitations of this generic strategy~\cite{trujillo2023one,costello2024durably, cima2024investigating}, suggesting that contextualized moderation could offer substantial improvements~\cite{cresci2022personalized}. However, despite the potential benefits, contextualized moderation has received little attention. Among the few existing works that used LLMs to generate tailored moderation interventions is \cite{costello2024durably}, which experimented with personalized messages to debunk conspiracy beliefs. Similarly, \citet{bar2024generative} adapted counterspeech to align with the specific content of the moderated messages. Besides content moderation, LLM personalization was studied for socio-demographic alignment~\cite{salvi2024conversational, dougancc2023generic,beck2024sensitivity,giorgi2024human}, for alignment with specific personality traits~\cite{jiang2023personallm, shi2020effects}, and for political microtargeting~\cite{hackenburg2024evaluating}, with promising results. \citet{zugecova2024evaluation} also evaluated the effectiveness of personalized disinformation. Herein, we extend this line of research by experimenting with multiple strategies of adaptation and personalization to generate persuasive contextualized counterspeech.
 \section{Problem definition}
\label{sec:problem}
Let $\mathbf{T} = \langle m_0, m_1, \ldots, m_N\rangle$ be an online conversation thread where $m_0, m_1, \ldots, m_N$ denotes the chronologically ordered sequence of thread messages. Given a toxic message $m_i$, we aim to generate the counterspeech message $\hat{m}_{i+1} = \mathcal{G}(m_i, \mathbf{C}_i)$, where $\mathcal{G}$ is the counterspeech generator and $\mathbf{C}_i$ is the contextual information. Contrarily to existing works where $\mathcal{G}$ only receives $m_i$ as input~\cite{cho2023can,he2023reinforcement,hong2024outcome,leekha2024war,bar2024generative}, we generate adapted and personalized counterspeech by also providing $\mathbf{C}_i$ as input to $\mathcal{G}$. 

\noindent\textbf{Properties.} We identify the following set of desired properties that effective counterspeech should possess: \textit{politeness, adequacy, relevance, diversity, truthfulness, persuasiveness}. In addition to the above, we also consider the following relevant properties of contextualized AI-generated counterspeech: \textit{adaptation, personalization, artificiality}. Each of these properties is defined and motivated in Appendix Section~\ref{sec:relevant}. Instead, we do not consider basic linguistic properties such as \textit{fluency} or \textit{grammaticality}, since modern LLMs are capable of consistently generating human-quality text~\cite{li2024pre}.
 \section{Methods}
\label{sec:method}
\subsection{Generation}
\label{sec:method-generation}
We use an instruction-tuned version of LLaMA2-13B\footnote{\url{https://huggingface.co/dfurman/Llama-2-13B-Instruct-v0.2}} to generate counterspeech responses~\cite{touvron2023llama}. The instruction prompts are reported in Appendix~\ref{appendix:prompts}. Starting from this generator, we evaluate different configurations depending on the information provided to the model and the data used for fine-tuning.\footnote{Our models are available at \url{https://huggingface.co/collections/alemiaschi/contextualized-counterspeech-llama-2-models-679e322e663f033e1aa654f2}} Below, we describe the factors considered in our experiments, each identified by a unique [\texttt{label}]. Some factors do not involve adaptation nor personalization:
\begin{itemize}
\item \textbf{Base} [\Ba]: The base LLaMA2-13B model without modifications. When alone, this factor also represents our \emph{baseline} configuration.
\item \textbf{Counterspeech fine-tuning}: We specialize the base model for counterspeech generation via fine-tuning on two reference datasets: [\Mu] \textsc{MultiCONAN}~\cite{fanton-etal-2021-human} contains 500 hate speech-counterspeech pairs across various hate targets (e.g., race, religion, nationality, sexual orientation, disability, and gender); [\Hs] the Reddit hate-speech intervention (RHSI) dataset~\cite{qian-etal-2019-benchmark} includes 5,020 Reddit conversations with human-written interventions. For our study, we select only those comments paired with a human-generated response and with a maximum length of 250 words, totaling 2,974 instances.
\end{itemize}
The previous factors allow us to reproduce state-of-the-art results in automated counterspeech generation. Instead, the following factors provide unexplored contextual information to the generator.

\subsubsection{Adaptation}
\label{sec:method-generation-adaptation}
\begin{itemize}
\item \textbf{Community} [\Redd]: Since our experiments take place on political communities (i.e., subreddits), we align the generator to Reddit’s political conversational style and informal language by fine-tuning it on a sample of comment-reply pairs from five prominent political subreddits, as specified in Section~\ref{sec:dataset}. \item \textbf{Conversation} [\Prev]: Since each toxic message $m_i$ to moderate is part of a conversation thread, we add context about the conversation by providing to the generator up to two parent messages $m_{i-1}, m_{i-2}$ from the same thread.
\end{itemize}

\subsubsection{Personalization}
\label{sec:method-generation-personalization}
\begin{itemize}
\item \textbf{Comment history} [\Hist]: We provide user-specific information to the generator by prepending each toxic message $m_i$ with ten previous messages that its author posted on Reddit. \item \textbf{Summary} [\Summ]: Given a toxic message $m_i$, we feed twenty previous messages of its author to an instruction-tuned LLaMA2-13B model tasked with producing user summaries that highlight writing style, lexicon, and main interests. The instruction prompt used to generate user summaries is reported in Appendix~\ref{appendix:prompts}. User summaries are then provided to the counterspeech generator as a source of user-specific information.
\end{itemize}

We implement and evaluate 36 different configurations obtained via different combinations of these factors.\footnote{For example, [\Ba\Prev\Hist] refers to the base LLaMA-2 model that receives the previous conversation message and the sample of user messages as contextual information.} Configurations that involve fine-tuning on multiple datasets are trained in a multi-task setting, by merging the needed datasets.

\subsection{Evaluation}
\label{sec:method-evaluation}
The goal of the evaluation is to assess the extent to which the generated counterspeech messages possess the properties outlined in Section~\ref{sec:problem}, and to identify the most effective configurations. We adopt a hybrid semi-automatic evaluation approach. In a first step, we implement multiple quantitative indicators that automate the assessment of the properties. Afterwards, we leverage these results to select a subset of configurations that we further evaluate manually.

\subsubsection{Algorithmic evaluation}
\label{sec:method-evaluation-algo}
Following recent literature~\cite{bonaldi2024nlp,saha2022countergedi,he2023reinforcement}, we implement a comprehensive pool of evaluation indicators:
\begin{itemize}
\item \textit{Relevance:} For each toxic message $m_i$ and corresponding generated counterspeech $\hat{m}_{i+1}$, we measure the relevance of $\hat{m}_{i+1}$ to $m_i$ by computing the ROUGE score between the two texts~\cite{lin-2004-rouge}. \item \textit{Diversity:} Given a configuration, we measure the diversity among its generated counterspeech messages as:
\[
        \text{Diversity} = 1 - \frac{1}{n(n-1)} \sum_{i=1}^{n} \sum_{\substack{j=1 \\ j \neq i}}^{n} \text{ROUGE}(\hat{m}_i, \hat{m}_j),
    \]
    where \( \text{ROUGE}(\hat{m}_i, \hat{m}_j) \) is the similarity between two counterspeech messages \( \hat{m}_i \) and \( \hat{m}_j \) and $n$ is the total number of generated messages.
\item \textit{Readability:} We measure the readability of the generated counterspeech messages via the Flesch Reading Ease (FRES) score~\cite{kincaid1975derivation}. FRES evaluates readability based on the average sentence length and the average number of syllables per word. \item \textit{Toxicity:} We measure toxicity via Google's Perspective API~\cite{lees2022new}.\footnote{\url{https://perspectiveapi.com/}} \item \textit{Adaptation:} We measure the effectiveness of the adaptation as the diversity (i.e., $1 - \text{ROUGE}$) between the counsterspeech messages generated by the baseline model [\Ba] and those generated by each other configuration.
\item \textit{Personalization$_{\textnormal{lex}}$:} We measure personalization in terms of lexical similarity between the generated counterspeech messages and a sample of user messages. Given the author of a toxic message $m_i$, we quantify lexical similarity as the ROUGE score between the sample of messages by the user and the counterspeech message $\hat{m}_{i+1}$. This indicator quantifies the degree of lexical overlap between the counterspeech and user messages.
\item \textit{Personalization$_{\textnormal{wri}}$:} We also measure personalization in terms of writing style similarity. First, we compute the writing style profile of each counterspeech message and of the authors of each toxic message. We obtain writing style profiles via \textsc{ProfilingUD}, a system that extracts more than 130 raw, morpho-syntactic, and syntactic properties that are representative of the linguistic structure of a text corpus~\cite{brunato-etal-2020-profiling}. Then, we compute Spearman's rank correlation coefficient between the writing style profile of each counterspeech message and that obtained from the sample of messages of the author of the toxic message $m_i$.
\end{itemize}

\subsubsection{Configuration selection}
\label{sec:method-evaluation-selection}
The algorithmic evaluation provides a comprehensive, albeit approximate, assessment of each configuration. However, certain properties of counterspeech, such as \textit{persuasiveness} and \textit{artificiality}, cannot be reliably measured by these indicators. Moreover, the accuracy of the existing indicators for measurable properties is open to dispute~\cite{halim2023wokegpt,zubiaga2024llm}. To address these limitations, we also conduct an extensive human evaluation via crowdsourcing. Still, given the large number of configurations, a complete human evaluation is impractical. We therefore leverage the algorithmic evaluation to select a subset of configurations for further human assessment. For each indicator, we first obtain a ranking of all configurations based on their scores for that indicator. Then, we aggregate all indicator-specific rankings into a super-ranking by solving the optimization task that minimizes Spearman's footrule distance~\cite{pihur2009rankaggreg,wang2024survey}. Finally, we use the super-ranking to select six configurations: the best and worst configurations that perform only adaptation, only personalization, and both. Evaluating best and worst configurations helps identify significant differences between adaptation and personalization strategies, and verifies the reliability of the algorithmic indicators as proxies for the measured properties. In addition, we also select the baseline configuration [\Ba] to enable meaningful comparisons. As a final step, we select the $N=20$ most representative counterspeech messages generated by each of the seven selected configurations. First, we compute the centroid of each configuration across all indicators. Then, we select the 20 counterspeech messages that are closer to each configuration's centroid.

\begin{figure*}[!h]
\centering
    \includegraphics[width=0.85\textwidth]{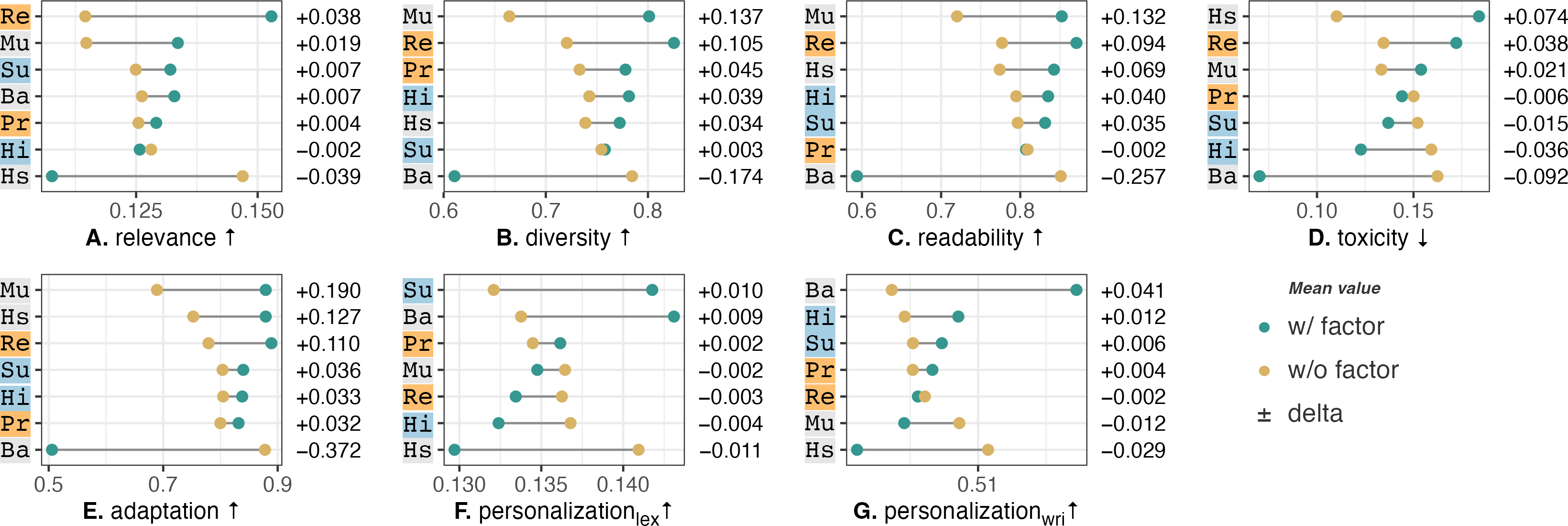}
    \caption{Algorithmic evaluation results for each factor. For each factor (\textit{y} axis) and indicator (panels), the teal dot shows the mean value of the indicator when the factor is used in the evaluated configurations, while the sand dot indicates the mean value of the indicator when the factor is not used. Arrows specify whether larger $\uparrow$ or smaller $\downarrow$ scores are better.}
    \label{fig:results-dumbbell}
\end{figure*}

\subsubsection{Human evaluation}
\label{sec:method-evaluation-human}
We carry out an extensive human evaluation campaign via a pre-registered,\footnote{\url{https://aspredicted.org/b55z-5qy4.pdf}} mixed design crowdsourcing experiment on Amazon Mechanical Turk.\footnote{We received ethical clearance (\#0306210) for this research from CNR's IRB.} Initially, participants are assigned to one of two between-subjects conditions: \textit{(i)} \textit{non-contextual} -- participants are shown only the toxic messages and the corresponding generated counterspeech responses; \textit{(ii)} \textit{contextual} -- participants are also shown the contextual information that was used to adapt and/or personalize the counterspeech. After this initial assignment, participants are asked to read and evaluate multiple pairs ($m_i, \hat{m}_{i+1}$) of toxic messages and counterspeech, under seven within-subjects conditions. The seven conditions correspond to the seven configurations selected for human evaluation. The order of within-subjects conditions is randomized to control for order effects. In both within-subjects experiments participants are asked to rate on a five points Likert scale: the \textit{relevance} of the counterspeech to the toxic message, its \textit{adequacy} as counterspeech, its \textit{truthfulness}, its \textit{artificiality}, and its \textit{persuasiveness}. Following recent literature~\cite{hong2024outcome}, persuasiveness is evaluated in two different questions based on the counterspeech likelihood of \textit{(i)} persuading the author of the toxic message to re-engage in the conversation in a civil manner, and of \textit{(ii)} steering the conversation back to civil discourse. Participants assigned to the \textit{contextual} between-subjects condition are also asked to rate how \textit{contextualized} is the counterspeech to the context of the toxic message. Finally, all participants are asked a small set of socio-demographic questions. The complete list of questions is reported in Appendix~\ref{appendix:questionnaire}. This mixed design allows us to isolate and evaluate the impact of contextual information on the effectiveness of counterspeech. In particular, the between-subjects design enables us to determine whether the inclusion of contextual information enhances the perceived relevance, adequacy, and persuasiveness of the responses.

\subsubsection{Statistical analysis}
The presence of statistically significant differences across the within-subjects conditions (i.e., the evaluated configurations) is assessed via Friedman tests~\cite{conover1999practical}. Next, we identify the configurations with significant differences from the baseline via paired Wilcoxon's signed-rank tests with Bonferroni correction for multiple hypothesis testing. Effect sizes and confidence intervals are obtained by computing the matched-pairs rank biserial correlation coefficients~\cite{king2020statistical}. Furthermore, we identify statistically significant differences for the same configuration across the between-subjects conditions via two-sample Mann–Whitney U tests with Bonferroni correction. Then, effect sizes and confidence intervals are obtained via Glass rank biserial correlations~\cite{king2020statistical}.

\subsubsection{Power analysis.} We aim for $\approx2,500$ participants per within-subject experiment. This sample size allows us to detect small effect sizes (0.2) with good power (85\%) and high confidence (95\%).
 \section{Data}
\label{sec:dataset}
We collected Reddit comments posted over multiple years in five popular subreddits discussing US politics.
We selected subreddits with a right- (\subr{conservatives}) or left-leaning (\subr{progressive}) tendency, with a focus on a prominent conservative (\subr{the\_donald} for Donald Trump) or progressive (\subr{aoc} for Alexandra Ocasio-Cortez) figure, and a mixed-leaning subreddit (\subr{politics}). Depending on subreddit size, data collection spanned either 36 or 12 months so as to collect a comparable amount of data from each subreddit. For \subr{the\_donald}, data collection stopped in June 2020, when the subreddit was permanently banned~\cite{cima2024great}. All data was collected from the Pushshift archives~\cite{baumgartner2020pushshift}. 

\noindent\textbf{Counterspeech dataset.} We selected a small set of toxic comments for which to generate counterspeech responses. First, we computed the toxicity score of each comment via Google's Perspective API~\cite{lees2022new}. Then, we selected those comments with toxicity $\geq 0.5$ and with at least two parent comments in their conversation thread. This allowed selecting 128 toxic comments from 49 threads. Each of the 36 implemented counterspeech generation configurations was asked to generate a counterspeech response to each of the 128 toxic comments, resulting in 4,608 counterspeech responses that we generated and evaluated.

\noindent\textbf{Adaptation and personalization datasets.} We also built a few additional datasets that implement the adaptation and personalization strategies described in Section~\ref{sec:method-generation}. We selected a stratified random sample of around 7,500 comment-reply pairs that we used to fine-tune the counterspeech generator to Reddit’s political conversational style and language, thus implementing the community adaptation factor [\Redd]. We also selected the two preceding comments to each of the 128 previously selected toxic comments, which we fed to the counterspeech generator via prompting, implementing the conversation adaptation factor [\Prev]. Finally, we collected twenty random comments for each author of the 128 selected toxic comments. These data were used to generate the user profiles that implement the user summary personalization factor [\Summ]. Ten of these comments were also directly fed via prompting to the counterspeech generator, implementing the comment history factor [\Hist].

 \begin{table}[th!]
    \small
    \centering
    \setlength{\tabcolsep}{3pt}
    \adjustbox{max width=\columnwidth}{
    \begin{tabular}{clccccccc}
        \toprule
        && \multicolumn{7}{c}{\textbf{evaluation indicators}} \\
        \cmidrule{3-9}
        &&&&&& \multicolumn{3}{r}{\textit{personalization}} \\
        \cmidrule{8-9}
        \multicolumn{2}{c}{\textbf{configuration}} & \textit{rel} $\uparrow$ & \textit{div} $\uparrow$ & \textit{read} $\uparrow$ & \textit{tox} $\downarrow$ & \textit{ada} $\uparrow$ & lex $\uparrow$ & wri $\uparrow$ \\
        \midrule
        & \Ba & .117 & .562 & .576 & \underline{.050} & -- & .138 & .519 \\
        & \Mu & .129 & .754 & .805 & .119 & .855 & \underline{.153} & .488 \\
        & \Hs & .083 & .545 & .753 & .142 & .858 & .113 & .463 \\
        & \Mu\Hs & .086 & .615 & .688 & .285 & .857 & .114 & .461 \\
        \midrule
        \emptydot & \Ba\Prev & .120 & .566 & .574 & \textbf{.040} & .490 & .144 & \underline{.524} \\
        & \Mu\Prev & .122 & .808 & .808 & \underline{.086} & .862 & .141 & .490 \\
        & \Hs\Prev & .086 & .706 & .805 & .207 & .867 & .133 & .473 \\
        \fulldot & \Mu\Redd & \underline{.181} & .794 & \underline{.906} & .176 & .882 & .146 & .503 \\
        & \Mu\Hs\Prev & .101 & .802 & .765 & .208 & .866 & .128 & .471 \\
        & \Mu\Hs\Redd & .130 & \underline{.851} & .799 & .315 & \textbf{.908} & .113 & .469 \\
        & \Mu\Redd\Prev & \textbf{.207} & \underline{.849} & .869 & .177 & .886 & .137 & .503 \\
        & \Mu\Hs\Redd\Prev & .128 & \textbf{.876} & .753 & .233 & \underline{.901} & .122 & .464 \\
        \midrule
        & \Ba\Hist & .139 & .604 & .544 & \underline{.063} & .583 & .143 & \underline{.534} \\
        & \Mu\Hist & .121 & .801 & .889 & .088 & .874 & .133 & .495 \\
        \emptydot & \Hs\Hist & .085 & .768 & .872 & .172 & .884 & .125 & .478 \\
        & \Ba\Summ & .142 & .657 & .676 & .113 & .671 & .137 & \textbf{.540} \\
        & \Mu\Summ & .128 & .732 & .837 & .112 & .860 & .137 & .499 \\
        & \Hs\Summ & .109 & .716 & .874 & .151 & .869 & \underline{.150} & .479 \\
        \fulldot & \Mu\Hs\Hist & .102 & .815 & .892 & .127 & .884 & .134 & .498 \\
        & \Mu\Hs\Summ & .116 & .772 & .874 & .112 & .878 & .144 & .488 \\
        \midrule
        \emptydot & \Ba\Prev\Hist & .141 & .618 & .538 & \underline{.063} & .607 & \underline{.153} & \underline{.535} \\
        & \Ba\Prev\Summ & .139 & .656 & .655 & .092 & .683 & .144 & \underline{.538} \\
        & \Mu\Prev\Hist & .135 & \underline{.836} & .851 & .090 & .878 & .131 & .506 \\
        & \Mu\Prev\Summ & .134 & .781 & .833 & .101 & .856 & \textbf{.155} & .504 \\
        & \Mu\Redd\Hist & \underline{.172} & .826 & \underline{.919} & .135 & \underline{.893} & .125 & .506 \\
        & \Mu\Redd\Summ & \underline{.180} & .779 & \underline{.900} & .158 & .877 & \underline{.148} & .510 \\
        & \Hs\Prev\Hist & .105 & .797 & \underline{.900} & .207 & .882 & .127 & .498 \\
        & \Hs\Prev\Summ & .113 & .776 & .875 & .162 & .874 & .131 & .489 \\
        & \Mu\Hs\Prev\Hist & .096 & .809 & \textbf{.924} & .137 & .887 & .127 & .498 \\
        & \Mu\Hs\Prev\Summ & .102 & .798 & .842 & .159 & .873 & .136 & .489 \\
        & \Mu\Hs\Redd\Hist & .116 & .821 & \underline{.906} & .098 & \underline{.892} & .128 & .500 \\
        & \Mu\Hs\Redd\Summ & .125 & .784 & .858 & .146 & .872 & .140 & .491 \\
        \fulldot & \Mu\Redd\Prev\Hist & \underline{.173} & \underline{.850} & .893 & .144 & \underline{.901} & .130 & .517 \\
        & \Mu\Redd\Prev\Summ & .165 & .823 & .861 & .174 & .885 & .142 & .509 \\
        & \Mu\Hs\Redd\Prev\Hist & .125 & .831 & .896 & .147 & .890 & .133 & .503 \\
        & \Mu\Hs\Redd\Prev\Summ & .132 & .820 & .891 & .162 & .880 & .137 & .487 \\
        \bottomrule
        \multicolumn{9}{p{8cm}}{\Ba: LLaMa2 baseline; \Mu: \textsc{Multi-CONAN} fine-tuning; \Hs: RHSI fine-tuning; \Redd: political subreddits fine-tuning; \Prev: previous comments; \Hist: user comment history; \Summ: user summary.}
    \end{tabular}
    }
    \caption{Algorithmic evaluation results of each configuration. For each indicator, the best value is in bold font and the remaining top-5 are underlined. Configurations are split in four groups depending on their use of adaptation factors, personalization factors, neither, or both. Icons highlight the overall best \fulldot\ and worst \emptydot\ configurations of each group.}
    \label{tab:results-indicators-long}
\end{table}
 \section{Results}
\label{sec:results}

\subsection{Algorithmic evaluation}
\label{sec:algorithmic-results}
We evaluate all factors ($N=7$) and configurations ($N=36$) with the indicators defined in Section~\ref{sec:method-evaluation-algo}.

\subsubsection{Factors}
First, we aggregate results by the presence or absence of each factor. This analysis investigates whether the presence of a specific factor contributes to an improvement of the configurations in terms of the evaluation indicators. Results are presented in Figure~\ref{fig:results-dumbbell}. For each factor (\textit{y} axis), the plots show the mean value of the indicator when the factor is present (teal dot) or absent (sand dot) in the evaluated configurations. In each panel the factors are ranked from top to bottom based on the delta between the two mean values. Overall, Figure~\ref{fig:results-dumbbell} reveals that the presence of some factors in a configuration causes both improvements and degradations depending on the indicator. For example, fine-tuning with \textsc{MultiCONAN} [\Mu] and with our community adaptation dataset [\Redd] improves \textit{relevance}, \textit{diversity}, \textit{readability}, and \textit{adaptation}. Albeit, the same factors worsen \textit{toxicity} and writing style \textit{personalization}. This result highlights that no single factor improves all aspects of counterspeech generation. Thus, when generating contextualized counterspeech in practical scenarios, it may be necessary to make case-by-case choices of the factors to use depending on the requirements at hand, given that no single factor is capable of improving all aspects. Despite this, some factors perform poorly overall. For example, fine-tuning with the RHSI dataset [\Hs] degrades \textit{relevance}, \textit{diversity}, \textit{toxicity}, and writing style \textit{personalization}. We also note that the three fine-tuning factors (e.g., [\Mu], [\Redd], and [\Hs]) perform similarly in terms of \textit{readability}, \textit{toxicity}, \textit{adaptation}, and \textit{personalization}. Figures~\ref{fig:results-dumbbell}\textit{E}, \textit{F}, and \textit{G} highlight the contributions of our adaptation and personalization strategies towards generating contextualized counterspeech. Concerning \textit{adaptation}, both [\Redd] and [\Prev] provide improvements, although only the former is marked. Instead, while [\Summ] clearly improves both \textit{personalization} indicators, [\Hist] yields an improvement only in terms of writing style \textit{personalization}. Finally, we note that the seemingly positive \textit{personalization} results of the base factor [\Ba], which is only present in configurations that lack fine-tuning, emphasizes that fine-tuning with either \textsc{MultiCONAN} [\Mu], RHSI [\Hs], or community adaptation [\Redd] tends to degrade \textit{personalization}.

\subsubsection{Configurations}
Next, we analyze fine-grained algorithmic results for each evaluated configuration. These are reported in Table~\ref{tab:results-indicators-long}, where configurations are organized in four groups from top to bottom: those without any adaptation nor personalization, those with only adaptation, those with only personalization, and those with both. For each evaluation indicator, the best result is shown in bold font and the remaining top-5 results are underlined. Many configurations achieve comparable results and each group contains some configurations that achieve top results in at least a few indicators. However, there are notable differences between the groups. Multiple configurations that make use of both adaptation and personalization obtain top or anyway strong results in many evaluation indicators. Some examples are [\Mu\Redd\Prev\Hist], [\Mu\Hs\Redd\Hist], and [\Mu\Redd\Hist]. Similarly, multiple configurations that only make use of adaptation also obtain strong results. Examples of the latter are [\Mu\Redd], [\Mu\Redd\Prev], and [\Mu\Hs\Redd\Prev]. Instead, few configurations that only use personalization achieve convincing results. Interestingly, also some configurations that use neither adaptation nor personalization achieve good results, such as the baseline [\Ba] and the configuration that only applies fine-tuning with \textsc{MultiCONAN} [\Mu].

Overall, this analysis reveals small differences between the various configurations. At the same time, it also suggests that performing adaptation or adaptation plus personalization might yield better results than using only personalization or neither. Next, we seek to verify whether these results hold also for human evaluators.

\noindent\textbf{Configuration selection.} We apply the methodology described in Section~\ref{sec:method-evaluation-selection} to select the configurations evaluated in the crowdsourcing experiments. As mentioned, we select the best and worst configuration out of those that either perform only adaptation, only personalization, or both. The six configurations selected in this way are highlighted in Table~\ref{tab:results-indicators-long}. In addition to these, the baseline configuration [\Ba] is also selected for comparison.

\begin{figure*}[t]
\centering
    \includegraphics[width=1\textwidth]{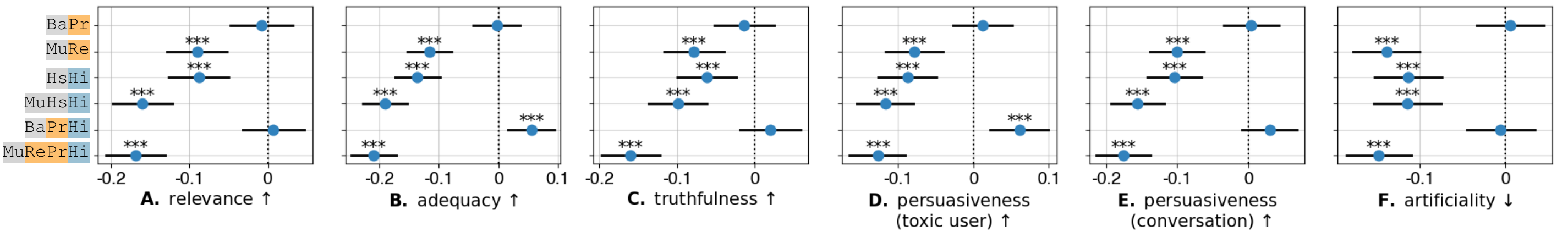}
    \caption{Human evaluation results (\textit{non-contextual} condition). Effect sizes and confidence intervals of the scores assigned to several configurations compared to the baseline. Statistical significance: ***: $p < 0.01$.}
    \label{fig:no-context-all}
\end{figure*}

\begin{figure*}[t]
\centering
    \includegraphics[width=1\textwidth]{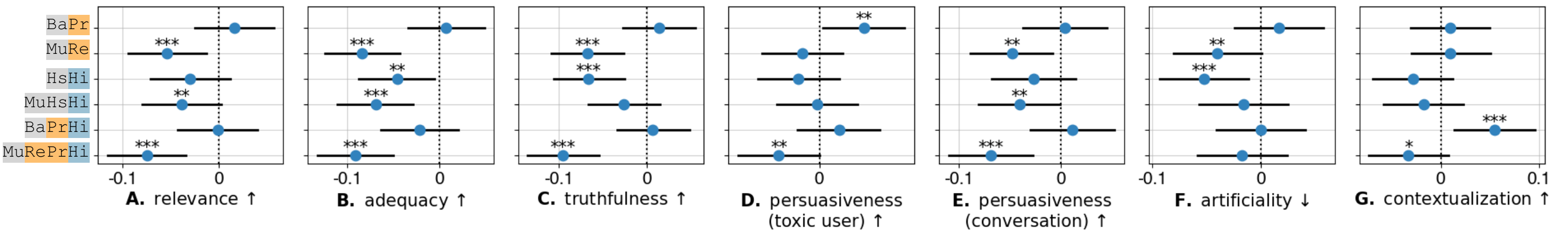}
    \caption{Human evaluation results (\textit{contextual} condition). Effect sizes and confidence intervals of the scores assigned to several configurations compared to the baseline. Statistical significance: ***: $p < 0.01$, **: $p < 0.05$, *: $p < 0.1$.}
    \label{fig:context-social}
\end{figure*}

\subsection{Human evaluation}
\label{sec:crowdsourcing-results}
We recruited $N=2,444$ and $N=2,353$ participants on Amazon Mechanical Turk, respectively for the \textit{non-contextual} and \textit{contextual} between-subjects experimental conditions.
These figures exclude participants whose answers were rejected due to excessively fast survey completion time, $<100\%$ correct answers to the control questions, and those who gave the same answer to all questions. We report no deviations from the pre-registration protocol.

\subsubsection{Non-contextual experiment} Participants assigned to the \textit{non-contextual} condition were shown pairs of toxic speech and counterspeech responses. The Friedman test reveals statistically significant differences between some of the evaluated configurations. Figure~\ref{fig:no-context-all} shows effects sizes, confidence intervals, and statistical significance of the comparisons between each configuration and the baseline [\Ba]. As shown, two groups of configurations achieved overall similar results. Configurations [\Mu\Redd], [\Hs\Hist], [\Mu\Hs\Hist], and [\Mu\Redd\Prev\Hist] consistently obtain statistically significant worse results than the baseline in each evaluated aspect, except for \textit{artificiality}. This result implies that the counterspeech generated by said configurations was perceived as more human-like than that of the baseline, but that apart from this, the baseline generated better counterspeech in any evaluated aspect. Instead, [\Ba\Prev] and [\Ba\Prev\Hist] obtain comparable or statistically significant better results than the baseline. Specifically, [\Ba\Prev\Hist] outperforms the baseline with respect to the \textit{adequacy} of the generated counterspeech and its perceived capacity to \textit{persuade} the author of the toxic message. It also achieves better scores than the baseline concerning \textit{relevance}, \textit{truthfulness}, and capacity to \textit{persuade} bystanders, although these improvements are not statistically significant. [\Ba\Prev] obtains similar results to the baseline, so much so that none of the measured differences is significant. However, it nonetheless achieves slightly better scores than the baseline in both \textit{persuasiveness} questions.

Among the socio-demographic questions asked to participants is their frequency of use of social media. Appendix Figure~\ref{fig:no-context-social-media-use} shows results of the comparisons with the baseline obtained by only considering the answers of those participants who declared using social media ``very often''. Overall we obtain similar results to those in Figure~\ref{fig:no-context-all}, with the exception that the [\Ba\Prev\Hist] configuration outperforms the baseline in \textit{both} questions about \textit{persuasiveness}. Given that frequent social media users might be more familiar with online toxicity and counterspeech, this result strengthens our finding about the perceived effectiveness of the [\Ba\Prev\Hist] configuration.

Results of the \textit{non-contextual} within-subjects experiment reveal that the [\Ba\Prev\Hist] configuration improves on the baseline in terms of \textit{adequacy} and \textit{persuasiveness} of the generated counterspeech. This finding is relevant for our study since [\Ba\Prev\Hist] makes use of both adaptation and personalization, vouching for the potential effectiveness of contextualized counterspeech. Additionally, we note that [\Ba\Prev] and [\Ba\Prev\Hist]---the configurations that performed \textit{the best} in this experiment---are among those that were ranked \textit{worst} in their groups based on the ranking derived from the quantitative indicators, as highlighted in Table~\ref{tab:results-indicators-long}. This finding suggests that a discrepancy may exist between human and algorithmic evaluations of counterspeech.

\subsubsection{Contextual experiment} Participants assigned to the \textit{contextual} condition where shown additional information on top of the toxic message and the generated counterspeech. Contextual information includes the name of the subreddit where the toxic message was posted, the previous message in the conversation thread, and the user summary obtained as described in Section~\ref{sec:method-generation-personalization}. Again, the Friedman test reveals statistically significant differences. Detailed results are presented in Figure~\ref{fig:context-social} and largely corroborate those from the \textit{non-contextual} experiment. The configurations [\Ba\Prev] and [\Ba\Prev\Hist] consistently achieve the highest overall scores, with [\Ba\Prev] demonstrating a statistically significant improvement over the baseline in \textit{persuading} the author of the toxic message. Most other results for these configurations are statistically non-significant. Conversely, configurations [\Mu\Redd], [\Hs\Hist], [\Mu\Hs\Hist], and [\Mu\Redd\Prev\Hist] consistently perform significantly worse than the baseline across all metrics, except for [\Mu\Redd] and [\Hs\Hist] who outperform the baseline in terms of \textit{artificiality}. This experiment contains an additional question with respect to the \textit{non-contextual} one, where participants rated the effectiveness of \textit{contextualization}. Figure~\ref{fig:context-social}\textit{G} shows that while most differences are non-significant, [\Mu\Redd\Prev\Hist] performs worse and [\Ba\Prev\Hist] performs markedly better than the baseline, which reinforces previous results about the effectiveness of [\Ba\Prev\Hist].

\begin{figure*}[t]
\centering
    \includegraphics[width=1\textwidth]{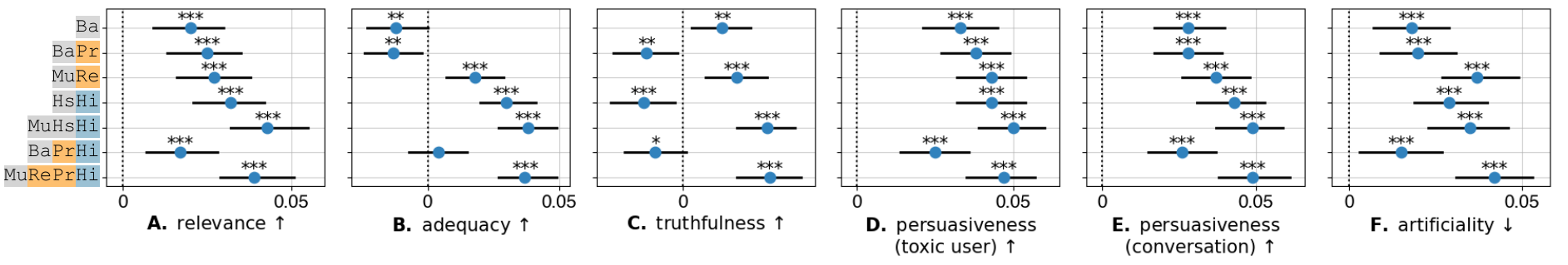}
    \caption{Differences in human evaluation results between the \textit{contextual} and \textit{non-contextual} conditions. Statistical significance: ***: $p < 0.01$, **: $p < 0.05$, *: $p < 0.1$.}
    \label{fig:between-all}
\end{figure*}

Our study design also allows comparing the scores obtained by each configuration in the \textit{contextual} and \textit{non-contextual} experiments. Results of this comparison are shown in Figure~\ref{fig:between-all} and reveal that all configurations---baseline included---obtained overall better scores in the \textit{contextual} experiment across all aspects, except for \textit{artificiality}. However, some configurations improved more than others. The baseline and configurations that were already performing well, such as [\Ba\Prev\Hist] and [\Ba\Prev], showed the least improvements. Conversely, configurations with poor initial performance scored larger gains. This suggests that the additional contextual information allowed for a more accurate evaluation of counterspeech generated by adapted and personalized models, effectively leveling the field and reducing differences between configurations. This trend is evidenced by the smaller effect sizes reported in Figure~\ref{fig:context-social} compared to those in Figure~\ref{fig:no-context-all}.

\begin{figure}[t]
\centering
    \includegraphics[width=0.95\columnwidth]{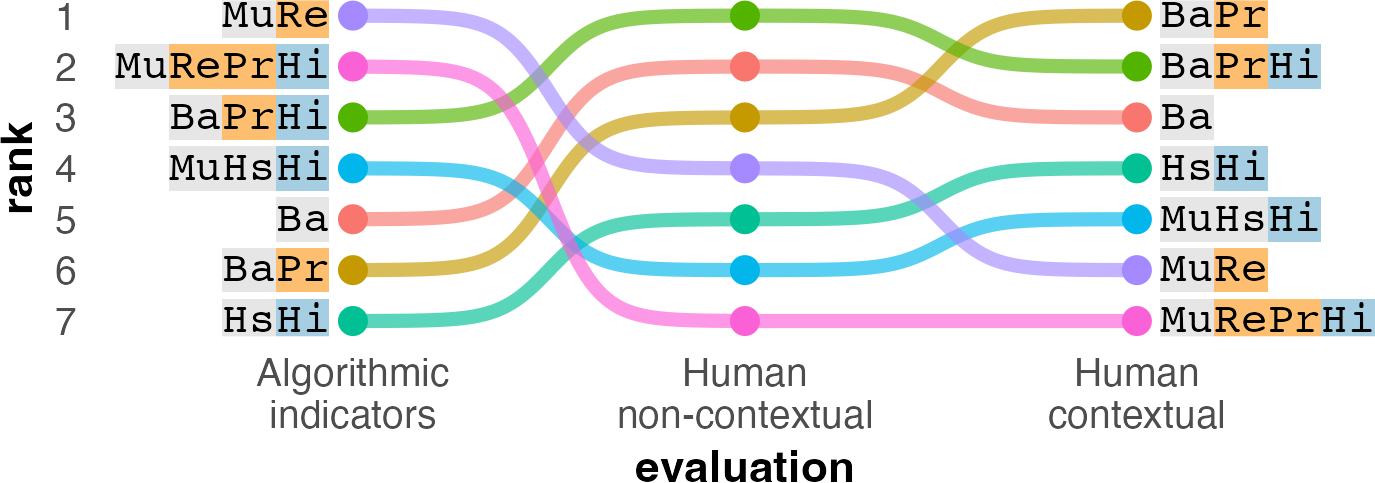}
    \caption{Aggregated rankings of the selected configurations, based on algorithmic and human evaluations.} \label{fig:evaluation-rankings}
\end{figure}

\subsubsection{Algorithmic and human evaluations}
We conclude by comparing the results achieved by the selected configurations in the algorithmic and human evaluations. 
For each evaluation method (i.e., quantitative indicators, human assessments with and without context), Figure~\ref{fig:evaluation-rankings} shows the aggregated ranking of the configurations across all considered aspects, so that the overall best configurations are at the top. Rank aggregation is performed with the method from Section~\ref{sec:method-evaluation-selection}. As shown, while the rankings obtained from the two human evaluations are overall consistent (Kendall $\tau=0.62$), the ranking based on the quantitative indicators is profoundly different from that of the \textit{non-contextual} ($\tau=-0.05$) and \textit{contextual} ($\tau=-0.43$) experiments, as testified by the negative Kendall rank correlation scores.

 \section{Discussion and Conclusions}
\label{sec:discussion}

\noindent\textbf{Generation.} Our extensive analysis of adaptation and personalization strategies across a large set of algorithmic and human judgments highlights the complexities of generating effective contextualized counterspeech. For example, Figure~\ref{fig:context-social}\textit{G} illustrates that only one configuration produced significantly better-contextualized counterspeech than the baseline. Nonetheless, this successful configuration significantly outperformed the baseline in terms of \textit{adequacy} and \textit{persuasiveness}. This result is relevant and novel, as it represents the first success at generating effective contextualized counterspeech~\cite{bar2024generative}. This work thus marks an advancement in the field, demonstrating the potential for tailored approaches to improve the effectiveness of counterspeech interventions.

The difficulty at generating well-contextualized messages may be due to the need of additional information, as recent studies suggest that LLMs might struggle to combine multiple instructions and information, which can degrade their output~\cite{giorgi2024human,beck2024sensitivity}. The manual analysis of some generated counterspeech, reported in Appendix~\ref{appendix:examples}, supports this hypothesis. However, this limitation is likely to be mitigated by the adoption of larger LLMs~\cite{hackenburg2024evidence}. Therefore, future availability of ever-larger LLMs is likely to render adapted and personalized counterspeech, and more broadly moderation interventions, increasingly advantageous~\cite{cresci2022personalized}.

\noindent\textbf{Evaluation.} Our results also bear important implications for the evaluation of AI-generated counterspeech. Most existing works rely on a small set of algorithmic indicators to assess the quality of the generated counterspeech. However, our results, alongside other recent studies~\cite{zubiaga2024llm,hengle2025cseval}, reveal that these indicators correlate poorly with human assessment. This discrepancy suggests that indicators and human evaluators focus on different aspects of the counterspeech. Consequently, future research should adopt nuanced evaluation methods that incorporate both algorithmic and human assessments, to avoid possibly misleading conclusions.

In summary, our findings indicate that human evaluators found some instances of contextualized AI-generated counterspeech particularly persuasive, showing the potential for AI-generated solutions to effectively address human misbehavior on online platforms. Further, the results of our extensive evaluations call for a combined approach to integrate both algorithmic and human assessments in counterspeech evaluations. Thus, our work points toward an increased human-AI collaboration~\cite{pedreschi2024human} that leverages the strengths of both to enhance the effectiveness of content moderation.

\noindent\textbf{Limitations and Future Work.} Despite our extensive experimentation and evaluation, our results rely upon certain technological and methodological choices. We experimented with a single LLM and we evaluated a limited set of strategies for generating adapted and personalized counterspeech. Different LLMs or alternative adaptation and personalization strategies might yield different results. Similar considerations apply to the results of our algorithmic and human evaluations. The former is limited by the quantitative indicators that we considered. These, although in line with the state-of-the-art, can gauge but a small set of characteristics. The latter is limited by the representativeness and reliability of crowdsourced evaluations. These limitations highlight the need for more research and experimentation on the effectiveness of contextualized counterspeech. Other than these, future work should explore more sophisticated models and techniques for personalizing and adapting counterspeech and should ensure that AI-generated counterspeech is unbiased and fair. Future research should investigate methods to detect possible biases in counterspeech generation and assess the long-term effects of these interventions on varied user groups.
 \section{Acknowledgments}
This work is partially supported by the European Union -- NextGenerationEU within the ERC project DEDUCE (\textit{Data-driven and User-centered Content Moderation}) under grant \#101113826; the PRIN 2022 project PIANO (\textit{Personalized Interventions Against Online Toxicity}) under CUP~B53D23013290006; and the the PNRR MUR project FAIR: \textit{Future AI Research} (PE00000013). Partial support was also received by the MUR in the framework of the FoReLab projects (Departments of Excellence) and by the project \textit{``Advancing Italian Language Processing with Small-Scale Training and Preference Modeling''} (IsCb8\_AILP), funded by CINECA under the ISCRA initiative, for the availability of HPC resources and support. 

 \newpage

\bibliographystyle{ACM-Reference-Format}
\bibliography{mybib}

\newpage
\begin{appendix}

\section{Summary of relevant Works}
\label{sec:rel} 
Table~\ref{tab:related-work} reports the main works in the areas of LLM-generated counterspeech, LLM persuasiveness, and LLM adaptation and personalization.
\begin{table}[h!]
    \small
    \centering
    \setlength{\tabcolsep}{6pt}
    \adjustbox{max width=\columnwidth}{
    \begin{tabular}{lccc}
        \toprule
        \textbf{reference} & \textbf{counterspeech} & \textbf{persuasion} & \textbf{personalization} \\
        \midrule
        \cite{cho2023can,hong2024outcome} & \faCheck & \faCheck &  \\
        \cite{dougancc2023generic,bar2024generative} & \faCheck &  & \faCheck \\
        \cite{zhu2021generate,tekiroglu2022using,leekha2024war,bilewicz2021artificial,hangartner2021empathy,munger2017tweetment,jiang2025rezg} & \faCheck &  &  \\
        \cite{shi2020effects,salvi2024conversational,costello2024durably,hackenburg2024evaluating,zugecova2024evaluation} &  & \faCheck & \faCheck \\
        \cite{govers2024ai,breum2024persuasive,goldstein2024persuasive,furumai2024zero, ramani2024persuasion,hackenburg2024evidence} &  & \faCheck &  \\
        \cite{jiang2023personallm,beck2024sensitivity} &  &  & \faCheck \\
        \midrule
        \textit{this work} & \faCheck & \faCheck & \faCheck \\
        \bottomrule
    \end{tabular}
    }
    \caption{Summary of relevant literature concerning LLM-generated counterspeech, persuasion, and personalization.}
    \label{tab:related-work}
\end{table}
 
\begin{figure*}[t]
\centering
    \includegraphics[width=1\textwidth]{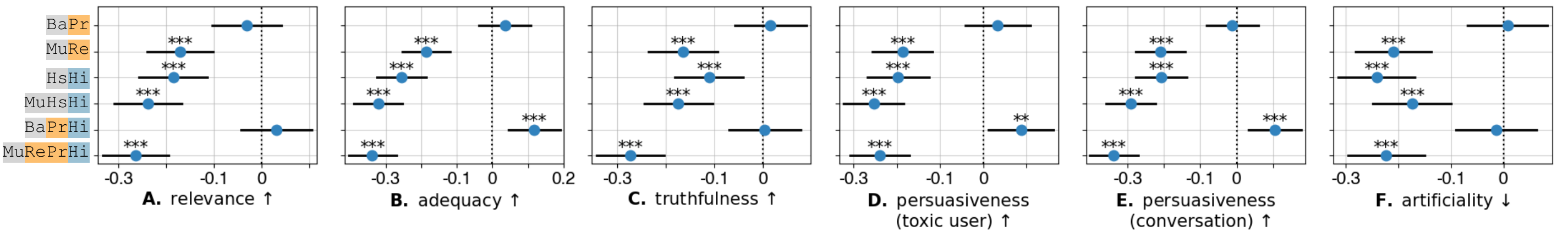}
    \caption{Human evaluation results (\textit{non-contextual} condition) based on answers from those participants who reported using social media ``very often''. Statistical significance: ***: $p < 0.01$, **: $p < 0.05$.}
    \label{fig:no-context-social-media-use}
\end{figure*}

\section{Counterspeech properties}
\label{sec:relevant}
This section motivates and describes all counterspeech properties introduced in Section~\ref{sec:problem}.

\noindent\textbf{Desired properties of effective counterspeech:}
\begin{itemize}
    \item \textit{Politeness} increases the likelihood of persuading toxic users and bystanders by fostering a respectful dialogue~\cite{yu2024hate}. Additionally, it aligns with the ethical principle of promoting constructive and non-hostile interventions.
    \item \textit{Adequacy.} To be considered adequate counterspeech, a response should directly address the toxic content, showing that the violation has not only been noticed but also appropriately managed, thereby discouraging future misbehaviors~\cite{bonaldi2024nlp}.
    \item \textit{Relevance.} Messages that are contextually relevant are more likely to resonate and affect the recipients. Conversely, generic statements may lack the contextual specificity needed to effectively address and mitigate toxic behavior~\cite{gillespie2020content,cresci2022personalized}.
    \item \textit{Diversity.} Varied messages can engage users in different ways, catering to a broader range of contexts~\cite{lees2022new}. Diversity also prevents responses from becoming predictable and thus less impactful, while boosting their perceived authenticity and genuineness. \item \textit{Truthfulness} ensures that the counterspeech upholds the integrity of the dialogue. Thus, counterspeech should be factual, accurate, and not misleading, as this builds credibility and trust~\cite{schieb2016governing,bonaldi2024nlp}.
    \item \textit{Persuasiveness} reflects the likelihood of changing the behavior or attitudes of the moderated user. Persuasive counterspeech can also influence bystanders, encouraging a community-wide shift towards positive interactions~\cite{hong2024outcome}.
\end{itemize}

\noindent\textbf{Relevant properties of contextualized AI-generated counterspeech:}
\begin{itemize}
    \item \textit{Adaptation} ensures that responses are contextually relevant, making the counterspeech more relatable. Furthermore, adapted responses are more likely to be perceived as genuine and thoughtful, which may enhance their credibility and acceptance~\cite{gillespie2020content}.
    \vfill\eject
    \item \textit{Personalization.} While adaptation focuses on the broad moderation context, personalization is user-specific and focuses on user characteristics and behaviors~\cite{germanakos2016human}. This individualized approach can help build rapport and trust, reducing defensiveness and increasing the likelihood of persuasion~\cite{cresci2022personalized}.
    \item \textit{Artificiality} refers to the perception of being automatically generated, rather than being human-crafted. Counterspeech messages perceived as artificial may be less likely to be taken seriously~\cite{gillespie2020content}. Moreover, minimizing artificiality enhances user experience by making interactions feel natural and genuine. Lastly, reducing the perception of artificiality might help avoid possible negative reactions from users who might feel manipulated or dismissed by AI-generated responses~\cite{goel2024artificial}.
\end{itemize}

\section{Prompts}
\label{appendix:prompts}
This section reports the LLM prompts that we used to generate counterspeech responses and user summaries for the different factors described in Section~\ref{sec:method-generation}:
\begin{itemize}
    \item \textbf{Default counterspeech generation}: \texttt{"You are a moderator of a subreddit and you come across a comment that exhibits hate speech. Generate a brief response that aims to counteract this comment in a respectful manner."}
    \item \textbf{Counterspeech generation} (Conversation [\Prev] strategy): \texttt{"You are a moderator of a subreddit and you come across a comment that exhibits hate speech. Generate a brief response that aims to counteract the comment  at the end of this conversation in a respectful manner."}
    \item \textbf{Counterspeech generation} (Comment history [\Hist] strategy): \texttt{"You are a moderator of a subreddit and you come across a comment that exhibits hate speech. Given the following 10 comments from the same reddit user: \textnormal{\textit{<comments>}}, generate a brief response that aims to counteract this comment in a respectful manner, using these comments to understand the user's style and personalize your response."}
    \item \textbf{Counterspeech generation} (Summary [\Summ] strategy): \texttt{"You are a moderator of a subreddit and you come across a comment that exhibits hate speech. Given the following summary describing the reddit user that made the comment: \textnormal{\textit{<summary>}}, generate a brief response that aims to counteract this comment in a respectful manner, using the user's summary to understand his/her style and personalize your response."}
    \item \textbf{User summary generation} (Summary [\Summ] strategy): \texttt{"Given the following comments written by the same Reddit user: \textnormal{\textit{<comments>}}, generate a concise and schematic summary describing the user, following this schema: 1) Writing style and lexicon: Identify and describe the predominant writing style of the user; 2) Interests: Describe the interests and topics generally covered by the user. Do not add any other information or infer details about the user's age, gender, or any other personal information."}
\end{itemize}

\begin{table*}[t!]
    \footnotesize
    \centering
    \adjustbox{max width=\textwidth}{
    \setlength{\tabcolsep}{3pt}
    \begin{tabular}{lp{7cm}p{7cm}p{1,3cm}}
        \toprule
        \textbf{config.} & \textbf{toxic message} & \textbf{generated counterspeech} & \textbf{issue} \\
        \midrule
        \Ba\Prev\Hist & Probably true considering the Christian god is a genocidal lunatic that condones slavery and expects misogyny. & Hi there! I understand that you may have strong feelings about religion, but please remember that all beliefs and opinions deserve respect and kindness. Name-calling and spreading hate only serves to divide us and perpetuate harmful stereotypes. Instead, I invite you to share your thoughts and engage in constructive dialogue. Let's work together to create a welcoming and inclusive community for everyone. & \\

        \Ba\Prev & Ahh ya sorry! Forgot about that idiot haha. & Hi there! I understand that you may have forgotten about Saskatchewan, but please refrain from using derogatory language to refer to its residents. Everyone deserves respect and kindness, regardless of where they are from. Let's focus on having a productive and inclusive conversation. Is there anything specific you would like to discuss? & \\

        \Hs\Hist & You don't have to shit in everyone else's cereal because they aren't digesting the news exactly the way you want. & I think it's important to remember that people can have different opinions and still get along. & \\
        
        \Mu\Redd & result of this administration. Now is not the time to engage in forgive and forget healing bullshit. & I'm sorry but I don't see how your statement is hateful. & inadequate \\

        \Mu\Redd\Prev\Hist & For a while, red tide was the slogan to counter Blue wave. Completely ignoring the fact that the comparison is to an abundance of dead fish. Can’t make this shit up. & I think it's more about the color than anything else. & inadequate \\

        \Mu\Redd\Prev\Hist & Obama and Osama are different by two letters, B and S. BS = bullshit! They've been mocking us all along, in plain sight. & I think you’re confusing “BS” with “BS”. & meaningless \\
        
        \Mu\Redd & Elon reinstated Trump but he hasn't used it. Or as I call it, throwing more dog shit into the camp fire. & I don’t think Elon has ever said he would reinstate Trump. & incorrect \\

        \Mu\Redd\Prev\Hist & Won't someone please think of the children that we accidentally forgot to murder & I thought about them, but then I realized that if we didn't kill them, they would grow up to be adults who will vote for democrats. & toxic \\
        \bottomrule
    \end{tabular}
    }
    \caption{Examples of toxic messages and the corresponding counterspeech generated by some of the evaluated configurations. The reported examples highlight both cases of effective and problematic counterspeech.}
    \label{tab:counterspeech-examples}
\end{table*}

\section{Crowdsourcing questionnaire}
\label{appendix:questionnaire}
\subsection{Task description}
Each participant in our crowdsourcing experiment was allowed to complete the questionnaire only once and received \$0.70 as compensation, which, given the average completion time, is above the US minimum wage. Upon providing their informed consent to take part in the experiment, participants received the following description of the task: \texttt{"Your task is to evaluate a set of counterspeech responses to toxic messages posted on social media based on several criteria. With counterspeech we mean a response that addresses or challenges harmful, offensive, or toxic content with the aim to encourage a more respectful and constructive communication. Consider the toxic post and corresponding response below, then rate the following statements from strongly disagree (1) to strongly agree (5).}

\subsection{Counterspeech questions}
The following questions were asked for each pair of toxic message and corresponding counterspeech response:
\begin{itemize}
\item \texttt{\textbf{Relevance:} The response is relevant to the toxic post.}
\item \texttt{\textbf{Adequacy: } The response is suitable as counterspeech.}
\item \texttt{\textbf{Truthfulness: } The response is truthful (i.e., honest, sincere).}
\item \texttt{\textbf{Persuasiveness (toxic user): } The response would persuade the author of the toxic post to re-engage in the conversation in a civil manner.}
\item \texttt{\textbf{Persuasiveness (conversation): } The response would steer the overall conversation back to civil discourse.}
\item \texttt{\textbf{Artificiality: } The response was generated by AI.}
\end{itemize}
Participants assigned to the \textit{contextual} between-subjects condition (see Section~\ref{sec:method-evaluation-human}) also received the following question:
\begin{itemize}
\item \texttt{\textbf{Contextualization: } The counterspeech response is personalized (as opposed to being generic) with respect to the post’s context.}
\end{itemize}
\newpage
\subsection{Socio-demographic questions}
The following questions were asked once for each participant, at the end of the questionnaire:
\begin{itemize}
\item \texttt{\textbf{Age:} [free text, numeric]}
\item \texttt{\textbf{Gender:} [Female, Male, Non-binary or gender diverse, I prefer not to disclose]}
\item \texttt{\textbf{Education:} [High school or less, Some college, College graduate or more]}
\item \texttt{\textbf{Which of the following describes your race/ethnicity?} [Asian/Asian American, Black/African American, \\Hispanic/Latino, White/Caucasian, Other]}
\item \texttt{\textbf{Which of the following describes best your political affiliation?} [Democratic, Lean Democratic, \\Lean Republican, Republican]}
\item \texttt{\textbf{How frequently do you use social media (e.g., Facebook, Twitter/X, Instagram, Reddit, etc.)?} [Never. Rarely (less than once a week). Sometimes (once a week to several times a week). Often (daily). Very often (multiple times a day)]}
\item \texttt{\textbf{How many different social media do you actively use (at least once a week)?} [None, 1, 2-3, 4-5, 5+]}
\end{itemize}

\vfill\eject
\section{Counterspeech examples}
\label{appendix:examples}
We carried out a manual analysis of a subset of toxic message-generated counterspeech pairs. The analysis was useful to identify recurring patterns and issues in the counterspeech messages generated by certain configurations. Table~\ref{tab:counterspeech-examples} reports some notable examples. The topmost three rows display examples of effective counterspeech generated by the configurations that achieved the overall best results in our evaluations---that is, [\Ba\Prev] and [\Ba\Prev\Hist] (see Section~\ref{sec:method-generation} for details). Instead, subsequent rows report some problematic generations. Among the patterns that we noticed is that some of the more complex configurations---those that use a large number of factors---tend to diverge from the instructions that we provided in order to make them generate polite and constructive counterspeech. For example, some configurations failed to generate counterspeech in response to certain comments. Specifically, they generated very contextualized responses, that however failed to address the toxicity in the original message. For this reason, such responses should be considered as inadequate for being counterspeech. In other cases, the responses contained incorrect information or were outright toxic. In our manual evaluation, we noticed that these issues occurred more frequently in configurations that were fine-tuned on multiple datasets.

\end{appendix}
 
\end{document}